\documentclass[11pt]{article}
\usepackage{graphicx,floatflt,amssymb}
\newcommand{\ba}{\begin{array}}
\newcommand{\ea}{\end{array}}
\newcommand{\bd}{\begin{displaymath}}
\newcommand{\ed}{\end{displaymath}}
\newcommand{\be}{\begin{equation}}
\newcommand{\ee}{\end{equation}}
\newcommand{\bea}{\begin{eqnarray}}
\newcommand{\eea}{\end{eqnarray}}

\def\ltap{\raisebox{-.4ex}{\rlap{$\sim$}} \raisebox{.4ex}{$<$}}
\def\gtap{\raisebox{-.4ex}{\rlap{$\sim$}} \raisebox{.4ex}{$>$}}

\newcommand{\dk}{d\kappa}

\def\a{\alpha}

\def\l{\lambda}

\def\p{\pi}

\begin{document}
\pagestyle{empty}
\begin{flushright}
\texttt{hep-ph/0211074}\\
SINP/TNP/02-30\\
LPT-Orsay 02-97\\
\end{flushright}

\vskip 100pt

\begin{center}
{\Large \bf Can radiative magnification of mixing angles
occur for two-zero neutrino mass matrix textures?} \\
\vspace*{1cm}
\renewcommand{\thefootnote}{\fnsymbol{footnote}}
{\large\sf Gautam Bhattacharyya $^{a,b, \!\!}$
\footnote{E-mail address: gb@theory.saha.ernet.in}},
{\sf Amitava Raychaudhuri $^{c,\!\!}$
\footnote{E-mail address: amitava@cubmb.ernet.in}},
{\sf Arunansu Sil $^{c,\!\!}$
\footnote{E-mail address: arunansu\_sil@yahoo.com}}
\vskip 10pt
$^{a)}${\small \it Saha Institute of Nuclear Physics, 1/AF Bidhan Nagar,
 Kolkata 700064, India}\\
$^{b)}${\small \it Laboratoire de Physique Th\'eorique,
Universit\'e de Paris XI, B\^atiment 210, 91405 Orsay Cedex,
France}\\
$^{c)}${\small \it Department of Physics, University of Calcutta, 92
Acharya Prafulla Chandra Road, Kolkata 700009, India}\\
\normalsize
\end{center}

\begin{abstract}
Neutrino Majorana masses and mixings can be generated from a
dimension-5 operator within the standard model particle content.
After a review of the mechanism of radiative enhancement of the
mixing angle in a two-neutrino case, we consider three-flavour
mass matrices of two-zero texture generated from such an operator
and investigate the possibility of implementing the mechanism
here.  We observe that  radiative magnification of only the
solar angle is consistent with oscillation data on masses and
mixings, and that too for nearly degenerate neutrinos, with two
of them having opposite CP parities, while for hierarchical
masses the mechanism does not work. In supersymmetry or in an
extra-dimensional scenario the above features are
qualitatively unchanged.

\vskip 5pt \noindent
\texttt{PACS Nos: 14.60.Pq, 11.10.Hi} \\
\texttt{Key Words:~Neutrino masses and mixings, Renormalization Group}
\end{abstract}
\vskip 20pt

\newpage

\pagestyle{plain}
\setcounter{page}{1}

\renewcommand{\thesection}{\Roman{section}}
\setcounter{footnote}{0}
\renewcommand{\thefootnote}{\arabic{footnote}}

\section{Introduction}
The mixing angles of quarks are experimentally known to be small.
It is a natural expectation, mainly boosted by the idea of quark-lepton
unification, that lepton mixings will be small too.  But, quite
contrary to this, recent neutrino experiments indicate \cite{3gen} that two
of the three mixing angles, namely the solar and the atmospheric ones,
are large \cite{sun,atm}, while the third mixing angle (CHOOZ) is small
\cite{chz}.  So, it is of interest to look for mechanisms which can
naturally explain the large mixing angles in the neutrino sector
without conflicting with the spirit of grand unification.
Renormalization Group (RG) evolution of neutrino masses and mixing
angles offers one such mechanism (\cite{babu1}--\cite{antu2}).

To implement this mechanism, the first step is to make an assumption
about the elements of the neutrino mass matrix at some scale. In
general, an arbitrary three flavour neutrino mass matrix involves nine
model-independent complex parameters. As emphasized in \cite{fgm},
from the results of the ongoing and foreseen experiments it is not
possible to fully determine this mass matrix. So, as the authors of
\cite{fgm} argued, scenarios with structural simplicity need to be
conjectured in which some elements of the mass matrix in the flavour
basis are identically zero.  The number of such zeros depends on the
symmetry of these so called `texture zero' mass matrices. They
observed that a mass matrix with more than two zeros is incompatible
with data, and seven out of the fifteen possible two-zero texture mass
matrices are consistent with experiments.  The authors of \cite{dproy}
have carried this investigation one step further by showing that only
three of the above seven mass matrices -- the ones which predict
hierarchical neutrino masses (normal or inverted) -- survive if one
takes the atmospheric mixing angle to be {\em exactly} maximal
($\pi/4$), while those structures which yield quasi-degenerate masses
are excluded. Calculation of Majorana-type CP violating phases
associated with two-zero textures has been presented in \cite{xing}.
It should also be mentioned that texture zero up and down quark mass
matrices have been successfully used in the past to relate the ratios
of quark masses to their mixing angles \cite{rrr}.

In view of this recent interest in texture two-zero structures, and
more so for their predictive properties, in this paper we investigate 
whether radiative magnification of neutrino mixing angles can occur in
such schemes. We will consider both quasi-degenerate and hierarchical
mass matrices at a high scale and examine whether
a significant running of one or more angles is consistent with data on
masses and mixings. This way we complement the analyses in \cite{fgm}
and \cite{dproy} to seek whether such textures can be
embedded in a unification framework.

Let us briefly summarize what is already known.  It has been
demonstrated in \cite{babu1,chan1} that starting from a tiny mixing
angle between {\em two} active neutrinos at some high energy scale,
one can achieve large mixing at low energy through RG evolution. The
analyses in \cite{casas,chan2} are of very general nature and they
contain explicit expressions for RG evolution of masses and mixing
angles.  A detailed discussion of how to promote the analysis from two
to three generations with a special reference to the existence of
fixed points of mixing angles at low energy is contained there.  The
existence of these fixed points has been shown by the authors of
\cite{balaji1} to lead to a stable atmospheric mixing angle close to
the maximum value. In \cite{balaji2}, it has been shown that starting
from small mixing angles at a high scale, radiative correction can
generate large atmospheric mixing at low scale while keeping the other
two angles small, thus leading to a small angle MSW solution.  The
authors of \cite{antusch} have shown that, in a see-saw model,
starting from a bimaximal mixing at the GUT scale, the solar angle is
driven by RG evolution to a smaller value at a low scale while the
other two angles do not move appreciably.  In \cite{antu2}, it has
been shown that starting from a very small solar mixing at the GUT
scale, the LMA solution can be reached by RG running.  It should be
noted that many of the above analyses discuss SM and supersymmetry
together in the same breath as, except for some numerical alteration,
the general parametrization remains the same for both scenarios.

The purpose of our analysis is two-fold. First, much new data
have accumulated, particularly from SNO. Bimaximal mixing is now
disfavoured, as is the SMA solution for solar neutrinos. Though
the best-fit atmospheric mixing has remained at the maximum value
over the years, solar mixing is now expected to be large but not
maximal.  Thus the conditions that have to be met through
radiative enhancement have changed. Second, we choose a
restricted framework, namely, the two-zero mass matrix textures.
With the aim of implementing the radiative enhancement mechanism
starting from small mixing at a high scale, we perform a
case-by-case consistency check with experimental data for a
variety of two-zero conjectures. We find that only those
textures which result in a  quasi-degenerate neutrino mass
spectrum support a large RG enhancement of the solar mixing
angle. We also briefly remark about scenarios beyond the SM,
namely, supersymmetry and extra dimensions.

To generate neutrino masses and mixings, we set $M_X$ as the
scale at which lepton number is broken.  The Majorana masses of the
left-handed neutrinos result from the following dimension-5
operator which involves only the SM fields:
\begin{equation}
L^{SM} = \frac{\kappa_{ij}}{M_X}\bar{\ell^c}_i\ell_j HH + {\rm h.c.}
\label{eq:base}
\end{equation}
where $\ell$ and $H$ are the left-handed lepton and Higgs doublets of
the SM, respectively.  Here $i,j$ are flavour indices. SU(2)
indices have been suppressed in (\ref{eq:base}). It results in a
neutrino mass matrix $M_{ij}\sim\kappa_{ij}(\upsilon^2/M_X)$, where
$\upsilon$ is the vacuum expectation value of the SM Higgs boson.

In section II, we write down the RG equation of $\kappa_{ij}$ and
review the two flavour formalism. In section III, we consider the
radiative corrections to three flavour two-zero mass
matrix textures classified in three different subsections based on the
hierarchy of mass patterns. We identify those where the radiative
enhancement mechanism can be implemented. In section IV, we carry the
discussion to supersymmetric and extra-dimensional scenarios. In
section V, we draw our conclusions. 

\section{Two flavour case}

Let us consider the following mixing matrix
\bea
U = \left[
          \begin{array}{cc}
          \cos\theta & \sin\theta  \\
          -\sin\theta & \cos\theta \end{array} \right],
\nonumber
\eea
such that $M = U M_d U^T$, where $M_d =$ diag $(m_1, m_2)$.
The neutrino mixing angle is then given by
\begin{equation}
\tan2\theta = \frac{2\kappa_{12}}{\kappa_{22}-\kappa_{11}} =
- \frac{2(\kappa_{12}/\kappa_{22})}{d\kappa},
\label{t2t}
\end{equation}
where we have defined
\[
d\kappa \equiv \frac{\kappa_{11}-\kappa_{22}}{\kappa_{22}}.
\]

The evolution of the $\kappa$ matrix is governed by the equation
\cite{babu1,chan1}\footnote{It was pointed out in \cite{an} that the
coefficient of the second term on the r.h.s. of Eq.~(\ref{rgk}) is
$-\frac{3}{2}$ in place of $-\frac{1}{2}$, commonly used in the earlier
literature.}
\bea
16\pi^2\frac{d\kappa}{d\ln\mu}=\{-3g_2^2+2\lambda+2S\}\kappa-\frac{3}{2}
\{\kappa({Y^\dag_l}{Y_l})+{({Y^\dag_l}{Y_l})^T}\kappa\},
\label{rgk}
\eea
where $\mu$ is the energy scale, $g_2$ is the SU(2) coupling constant,
$Y_l$ is the Yukawa coupling matrix of charged
leptons\footnote{We work in a basis in which $Y_l$ is diagonal.}, and
\[
S = {\rm Tr}~[3{Y_u}^\dag{Y_u} + 3{Y_d}^\dag{Y_d} +{Y_l}^\dag{Y_l}].
\]
Above, $Y_{u,d}$ are the Yukawa coupling matrices of the up- and
down-type quarks, respectively.
The RG equations for the SM couplings are well known
\cite{mtc} and are not presented. Here, the gauge and
Yukawa couplings at the electroweak scale are chosen as input
parameters.   It is necessary to exercise some caution in the
choice of the Higgs mass to fix the quartic scalar coupling
$\lambda$. Although the mixing angle evolution is quite
insensitive to this choice,  unless $m_H ~\gtap ~150$ GeV, the RG
evolution drives $\lambda$  to negative values  making
the scalar potential unbounded from below.

The first term in braces on the r.h.s. of (\ref{rgk}), which
we denote by $D \equiv -3g_2^2+2\lambda+2S$, treats all elements
of the $\kappa$ matrix identically -- a universal contribution --
while the term in the second braces involving the leptonic Yukawa
matrices distinguishes between different elements. Further, the
equation is linear in $\kappa$ so that the net effect of the RG
evolution is a multiplicative change for each element. More
specifically,

\bea
16\pi^2\frac{d\kappa_{11}}{d\ln\mu} & = & D \kappa_{11} -\frac{3}{2}
(2Y_1^2) \kappa_{11}, \nonumber \\
16\pi^2\frac{d\kappa_{12}}{d\ln\mu} & = & D \kappa_{12} -\frac{3}{2}
(Y_1^2 +Y_2^2) \kappa_{12}, \\
16\pi^2\frac{d\kappa_{22}}{d\ln\mu} & = & D \kappa_{22} -\frac{3}{2}
(2Y_2^2) \kappa_{22}. \nonumber
\eea
For our subsequent discussions, we take $Y_2 = Y_\tau$ and $Y_1 = Y_e$
or $Y_\mu$. Now, to a good approximation, the effects of RG
running can be  summarized by introducing two parameters $r$ and $a$
as:
\begin{eqnarray}
\label{ar}
\kappa_{11}(M_X) &\rightarrow& \kappa_{11}(M_Z) = a \kappa_{11}(M_X),
\nonumber \\
\kappa_{12}(M_X) &\rightarrow& \kappa_{12}(M_Z) =
a(1 + r/2) \kappa_{12}(M_X),
\\ \nonumber
\kappa_{22}(M_X) &\rightarrow& \kappa_{22}(M_Z) = a(1 + r)
\kappa_{22}(M_X),
\nonumber
\end{eqnarray}
where $M_X$ is some high scale and $M_Z$ characterizes the
electroweak scale. The universal contribution, $a \sim 0.7$ for
$M_X = 10^{18} - 10^{19}$ GeV, and is dominant. The other piece,
namely, $r \simeq (3Y_\tau^2/16\pi^2) \ln(M_X/M_Z) \sim Y_\tau^2
\sim 10^{-4}$ is crucial in determining the running of the mixing
angle. By considering a simultaneous one loop running of all the
necessary couplings, we have numerically checked that the
parametrization of (\ref{ar}), which we use only for
illustrative purposes, works extremely well to order $r^2 \sim
10^{-8}$. We should note that the value of $r$ is really
controlled by the $\tau$-lepton Yukawa coupling, and its
order-of-magnitude does not vary appreciably if $M_X$ is altered
even by a few orders.

The following points, based on which our subsequent
arguments will proceed, are worth noticing: \\
a) It follows from (\ref{ar}) that the quantity $\dk
\tan2\theta = - 2\kappa_{12}/\kappa_{22}$ is renormalization scale
invariant to a good approximation. \\
b) It also follows from (\ref{ar}) that
\[
\dk(M_Z) \simeq \dk(M_X) - r.
\]
c) Combining a) and b) we obtain
\bea
\tan 2\theta(M_Z) = \tan 2\theta(M_X)/\Delta, ~~{\rm where}~~
\Delta \equiv 1 - r/\dk(M_X).
\label{delta}
\eea

Now let us pay attention to the case when a small mixing angle at
$M_X$ becomes large near $M_Z$ by a resonant enhancement. Since
both the solar and atmospheric neutrino data suggest large
neutrino mixing, it is of more interest to consider this class of
running  than other possibilities.  Eq.~(\ref{delta}) indicates
that a resonant enhancement will be possible if $\dk(M_X)$ is
chosen to be as close to $r$ as possible. The quantity $\Delta$
is a fine-tuning parameter which determines this closeness.
Smaller the value of $\Delta$, more fine-tuned is the initial
texture; but paying this price we attain an enhancement of the
mixing by a factor $1/\Delta$, which is the ratio of the two
tangents as shown in the first equality of (\ref{delta}).  It
also turns out that
\bea
\dk(M_Z) = \Delta . \dk(M_X) \simeq \Delta . r \sim 10^{-4} . \Delta,
\label{dkmz1}
\eea
to ensure a significant running ending up in  large mixing near
the electroweak scale.

Let us now estimate the impact of (\ref{dkmz1}) on the neutrino masses
and mass splittings. For the sake of illustration, assuming $m_1 \sim
m_2 \gg |m_1 - m_2|$, we can express
\bea
\dk(M_Z) \simeq 0.5 \frac{\Delta m^2}{m^2} \cos2\theta(M_Z),
\label{dkmz2}
\eea
where $m = 0.5(m_1 + m_2)$ is an average mass, and $\Delta m^2 \equiv
|m_1^2 - m_2^2|$, both defined at the low scale.
Combining Eqs.~(\ref{dkmz1}) and (\ref{dkmz2}), we can write,
\bea
\Delta m^2 ({\rm eV}^2)~
\ltap~ 10^{-3} \left[\frac{\Delta}{\cos2\theta(M_Z)}\right]
\left[\frac{m}{2.2~{\rm eV}}\right]^2,
\label{master}
\eea
where the inequality arises from the upper limit on the absolute
neutrino masses (see discussions later).

The relation (\ref{master}) is fairly general. It only assumes
that $\nu_\tau$ has to participate in the oscillation, and that
there is a significant running of the mixing angle.
Quantitatively, the latter signifies that one can substitute
$\dk(M_X)$ by $r$ in (\ref{dkmz1}). Now we have to make a
judgement of what is the maximum value of $m$ that we are allowed
to take and how much fine-tuning, parametrized by the quantity
$\Delta$, we can tolerate.

The mass matrix that we have generated is of Majorana nature. The
neutrinoless double beta decay constraint \cite{db} applies on the
($ee$)-element of the mass matrix, and considering large mixing at low
scale, this means 0.05 $< m <$ 0.84 eV at 95\% C.L. But this result
needs further confirmation. Hence, a more conservative constraint is
to use $m < 2.2$ eV from the Tritium beta decay experiment \cite{tri}.

The size of $\Delta$ is indeed a tricky issue.  It follows from
(\ref{delta}) that $\Delta/\cos2\theta(M_Z) \simeq \tan2\theta(M_X)$
as $\theta(M_Z)$ gets closer to $\pi/4$.  To get an intuitive feeling
of the size of this fine-tuning, let us consider a toy scenario in
which a small angle $\alpha \equiv \theta(M_X)$ becomes $(\pi/4 -
\alpha) = \theta(M_Z)$ by a resonant enhancement. Then $\tan
2\theta(M_X) = \cot 2\theta(M_Z) \simeq \cos 2\theta(M_Z) \leq 0.22$
(post-SNO fit) \cite{fogli}. In this example, $\Delta \sim \cos^2
2\theta(M_Z) \leq 0.04$.  All in all, it is perhaps not unfair to take
$\tan2\theta(M_X) \sim 0.1 - 0.5$, which corresponds to $\theta(M_X)
\sim (3^\circ - 13^\circ)$ to cover a rather wide range of what we can
call a `small' initial mixing. Plugging this input in (\ref{master}),
we observe that $\Delta m^2 \leq 5 \times 10^{-4} ~{\rm eV}^2$ is a
very safe prediction (assuming $m \leq 2.2$ eV).

Now, if we attempt to explain the solar neutrino problem with the
above formalism, it is possible to arrange for both the LMA ($\Delta
m^2 \sim 10^{-5}$ eV$^2$) and LOW ($\Delta m^2 \sim 10^{-7}$ eV$^2$)
solutions. However, if the neutrinoless double beta decay observation
is confirmed, the situation will become rather tight.  For atmospheric
neutrinos, the preferred value of $\Delta{m^2}$ is $\sim 3\times
10^{-3}$ eV$^2$. It is obvious from (\ref{master}) that this cannot be
explained if we have to respect the Tritium beta decay limit on the
absolute neutrino mass.

We conclude this section by noting that instead of involving
$\nu_{\tau}$ if we consider $\nu_{\mu} - \nu_e$ oscillation, then, for
resonance, $\dk$ would have to be $\sim {Y_{\mu}}^2$ for which the
corresponding $\Delta{m^2}$ is too small to fit the experimental data.

\section{Three flavour case with two-zero mass matrix textures}
Now we turn to the question whether it is possible to accommodate both
the solar and atmospheric neutrino solutions in a picture of radiative
enhancement of mixing within a three flavour scenario, keeping in mind
the constraint from the CHOOZ experiment. We consider a solution
acceptable if both the mass splittings as well as the {\em
approximately} bimaximal nature of the mixing matrix are reproduced at
low energy, {\em with the radiative effects playing a significant
role}. The analysis in the case of the most general $(3 \times 3)$
structure is not very tractable. Some discussions are available in the
literature (\cite{casas}-\cite{antusch}). We restrict ourselves to the
two-zero texture mass matrices which have recently attracted
attention.  These matrices are defined in terms of six real parameters
and are written in a basis in which the charged lepton mass matrix is
diagonal.  We show that for these textures the conclusions can be
drawn in a rather simple and instructive fashion without taking
recourse to numerical calculations.

Two-zero $(3 \times 3)$ mass matrix textures of three different types
($A, B,$ and $C$) have been shown to be compatible with the present
experimental data \cite{fgm}. These textures correspond to
hierarchical, quasi-degenerate, and inverted hierarchical neutrino
masses, respectively \cite{dproy}. Since zeros of $\kappa_{ij}$ are
unaffected by RG evolution, the pattern of these neutrino mass
matrices -- i.e., the zeros in the texture -- is unaltered at the high
scale. The question we investigate here is whether RG evolution can
significantly affect the mixing angles for these mass matrix
textures. We consider the three cases in turn.

\subsection{Quasi-degenerate masses}
If at a high energy the neutrino mass matrix is of the
form
\bea
M_\nu^h = \left[
                 \begin{array}{ccc}
                 x_1 & x_2 & 0  \\
                 x_2 & 0 & x_3 \\
                 0 & x_3 & x_4
            \end{array} \right],
\label{b0}
\eea
then, in view of our discussions in the previous section, see
(\ref{ar}),  at the low scale it becomes
\bea
M_\nu^l = a \left[
                 \begin{array}{ccc}
                 x_1 & x_2 & 0  \\
                 x_2 & 0 & {x_3}(1+\frac{r}{2}) \\
                 0 & {x_3}(1+\frac{r}{2}) & {x_4}(1+r)
            \end{array} \right].
\nonumber
\eea

Notice that the overall scale factor $a \sim {\cal O}(1)$ does not
affect the mixing; only the mass eigenvalues are scaled by it.
The above mass matrix belongs to the type $B_1$ texture in the
notation of \cite{fgm}. It corresponds to neutrinos with nearly
degenerate masses \cite{fgm,dproy}.

Our goal will be to extract the mass splittings and mixing angles which
follow from this low energy neutrino mass matrix.  The 3 $\times$ 3
matrix $M_\nu$ can be diagonalized according to
\[
V^T M_\nu V = {\rm{diag}}(m_1,m_2,m_3),
\]
where $V$ is the Maki-Nakagawa-Sakata \cite{mns} unitary matrix which
relates flavour ($\a$) and mass ($i$) eigenstates of neutrinos through
$\nu_\a = V_{{\a}i}\nu_i$ and $m_1, m_2, m_3$ are the eigenvalues. It
can be expressed as $V = U_{23}U_{13}U_{12}$, where $U_{ij}$ are the
standard rotation matrices.

The rotation angle $\theta_{23}$ is given by
\[
\tan2\theta_{23}^l = \frac{2x_3(1+\frac{r}{2})}{x_4(1+r)}.
\]
It is obvious that there is no scope of getting large $\tan
2\theta_{23}$ at the low scale starting from a small one at the
high scale through a resonance induced mechanism. So, for this
texture, we have to keep $\tan 2\theta_{23}$ large for the whole
scale of running and the condition for this is\footnote{Here, and
in the following, $\tan 2\theta_{ij}$, ($i,j = 1,2,3$), are allowed
to be positive as well as negative (the so-called
`dark side').}
\be
\frac{|x_3|}{|x_4|} \gg 1.
\label{b2}
\ee

Applying the rotation through $U_{23}$, we get
\bea
{M_\nu^l}_{23} = a \left[
                 \begin{array}{ccc}
                 x_1 & x_2 c^l_{23} & x_2 s^l_{23}  \\
                 x_2 c^l_{23} & \lambda_1 & 0 \\
                 x_2 s^l_{23} & 0 & \lambda_2
            \end{array} \right].
\label{m23}
%\nonumber
\eea
where $s^l_{ij}$ and $c^l_{ij}$ are the sines and cosines of
$\theta^l_{ij}$. Also,
\be
\lambda_1 = x_4 s^{l2}_{23}(1+r) -
2x_3s^l_{23}c^l_{23}(1+\frac{r}{2}) \simeq  - x_3(1+\frac{r}{2}),
\label{blam1}
\ee
and
\be
\lambda_2 = x_4 c^{l2}_{23}(1+r) +
2x_3s^l_{23}c^l_{23}(1+\frac{r}{2}) \simeq  x_3(1+\frac{r}{2}).
\label{blam2}
\ee
where in the final step we have set $\theta_{23} = \pi/4$.

For the next rotation $U_{13}$, we set
\[
\tan2\theta^l_{13} = \frac{2x_2s^l_{23}}{\lambda_2 - x_1}
\simeq \frac{\sqrt{2}x_2}{x_3(1+\frac{r}{2}) - x_1}.
\]
The bound on ($V^l)_{e3}$ from CHOOZ requires the angle $\theta_{13}^l
\rightarrow$ 0. A significant RG evolution of this angle will demand a
large $\tan 2\theta^h_{13}$, which is possible if
\be
x_3 \simeq x_1.
\label{r1}
\ee
This, together with $\tan 2\theta^l_{13} \rightarrow 0$, and the
smallness of $r$ ($\sim 10^{-4}$), fixes the following relation
between mass matrix elements
\be
 |x_2|, |x_4| \ll |x_3| \sim |x_1|.
\label{r2}
\ee
Now, after the second rotation\footnote{The smallness of $\theta_{13}$
is a consequence of the (13)- and (31)-elements of (\ref{m23}) being
negligible compared to the (11)- and (33)-elements. The second rotation
therefore amounts to simply dropping the former elements. This
approximation is also used in the following subsections.}

\bea
{M_\nu^l}_{23,13} \simeq a \left[
                         \begin{array}{ccc}
                         x_1 & x_2/\sqrt{2} & 0  \\
                         x_2/\sqrt{2} & \lambda_1 & 0 \\
                         0 & 0 & \lambda_2
            \end{array} \right].
\nonumber
\eea
The next step is to diagonalize the (12)-block through the
$\theta_{12}$ rotation:
\[
\tan2\theta^l_{12} = \frac{\sqrt{2}x_2}{\lambda_1 - x_1}.
\]
It is obvious from (\ref{blam1}), (\ref{r1}), and (\ref{r2}) that
$\tan2\theta^l_{12}$ is small; a direct contradiction of the
empirical requirement.

Now we can check the only remaining possibility, i.e., keeping $\tan
2\theta_{13}$ {\em small} while, as before, $\tan 2\theta_{23}$
remaining large throughout the energy range, whether there can be a
prominent RG evolution of $\theta_{12}$. In this case, we get the
relation
\[
|x_3 - x_1| \gg \sqrt{2}|x_2|.
\]
It is seen that
\be
\tan2\theta^h_{12} \simeq \frac{\sqrt{2}x_2}{- x_3 - x_1}, \;\;\;
\tan2\theta^l_{12} \simeq \frac{\sqrt{2}x_2}{ - x_3
(1+\frac{r}{2}) - x_1}.
\label{b5}
\ee
A resonant enhancement of
$\tan2\theta_{12}$ requires $r x_3 \simeq -2(x_3 + x_1)$. Thus,
\be
x_3 \sim -x_1
\label{b6}
\ee
and, further, for significant running of
$\theta_{12}$, from (\ref{b5})
\be
r x_3 \gg 2 \sqrt{2} |x_2|.
\label{b7}
\ee
It now remains to verify whether it is possible to satisfy
(\ref{b2}), (\ref{b6}) and (\ref{b7}) and at the same time
reproduce the correct mass splittings at low energies for
atmospheric and solar neutrino oscillations. For this texture,
the mass eigenstates are quasi-degenerate\footnote{$m_1 \simeq
m_2 \simeq -m_3$; the negative mass eigenvalue representing
opposite CP-phase.} and it can be shown that:
\be
\Delta m^2_{atm} \sim 2 a^2 x_1 x_4, \;\;\; 
\Delta m^2_{sol} \sim 2 \surd2
a^2 x_1 x_2.
\label{b8}
\ee
Experiment demands that $\Delta m^2_{atm}/\Delta m^2_{sol} \sim 100$
for the LMA solution to the solar neutrino problem while it is $\sim
10^4$ for the LOW solution. From (\ref{b8}) it is clear that there is
no obstruction in achieving the desired ratio by suitably choosing
$x_2, x_3,$ and $x_4$.  Since the relationships amongst the $x_i$ are
linear, they can all be scaled to achieve the right absolute magnitudes
of the mass splittings.  

We have numerically checked that this is true.  For example, with the
inputs $x_1 = -1.799196, x_2 = 9.0 \times 10^{-6}, x_3 = 1.8$ and $x_4
= 1.8 \times 10^{-3}$ (all in eV), we observe a running of the solar
mixing angle and the mass splittings as given in Table
1\footnote{Variations of $x_i$ around the quoted values are tightly
constrained by the experimental data. Interestingly, the effective
Majorana mass parameter relevant for neutrinoless double beta decay is
predicted in the 1 eV range. We have also checked that both the
Tritium beta decay bound (mentioned earlier) and the cosmological
bound from the recent 2dF Galaxy Redshift Survey, namely $\sum_i
m_i~\ltap~2.2$ eV \cite{galaxy}, are satisfied.}.

\begin{table}
\begin{center}
\begin{tabular}{|c|c|c|c|c|c|}\hline
Scale&$\tan \theta_{23}$&$\sin 2\theta_{13}$&$\sin
2\theta_{12}$&$\Delta m^2_{atm}$ (eV$^2$)&$\Delta m^2_{sol}$ (eV$^2$)
\\ \hline
High & 0.9995 & 0.0 & 0.132 & 6.5$\times 10^{-3}$ & 3.5$\times
10^{-4}$ \\ \hline
Low  & 0.9995 & 0.0 & 0.907 & 3.2$\times 10^{-3}$ & 2.5$\times
10^{-5}$ \\ \hline
\end{tabular}
\begin{description}
\item{Table 1:} The mixing angles and mass splittings from the
chosen values of $x_i$ (see text).
\end{description}
%\caption{\sf \small The mixing angles and mass splittings
%from the chosen values of $x_i$ (see text).}
\end{center}
\end{table}

The fine-tuning between $x_1$ and $x_3$ above is an essential
ingredient of the radiative enhancement of the $\theta_{12}$ mixing
angle. This is reminiscent of the fine-tuning in the two-flavour case
discussed in the previous section. Note that $\tan \theta_{23}$ is
large but not maximal, a consequence of the non-zero value of $x_4$ --
a feature noted in \cite{dproy}. The quasi-degenerate neutrino
spectrum with one neutrino with an opposite CP-phase from the other
two is necessary for an RG evolution of $\theta_{12}$. This is
consistent with the observations in \cite{balaji3}.

There are three other two-zero texture mass matrices ($B_i$, $i$ =
2,3,4, in the notation of \cite{fgm}) for which RG running of mixing
angles can be significant as in the example discussed above. The $B_3$
texture differs from Eq.~(\ref{b0}) in that the (13)- and
(31)-elements are non-zero while the (12)- and (21)-elements
vanish. Obviously, this does not affect the $\theta_{23}$ prediction,
which is again maximal.  After the $\theta_{23}$ rotation through
$(\pi/4)$ the second and third rows (and columns) become the same in
the $B_1$ and the $B_3$ textures and the remaining discussions are
identical. Finally, the $B_2$ ($B_4$) texture can be obtained from the
$B_3$ ($B_1$) texture by placing the zero diagonal entry in the (33)
position rather than in the (22) position. This only affects the first
step of the argument in that $\tan 2\theta_{23}$ now changes
sign. Obviously this does not affect the conclusions in any way. {\em
Thus the four two-zero textures $B_i$, ($i = 1, \ldots, 4$), which
correspond to a quasi-degenerate neutrino mass spectrum, can all have
prominent RG running of the solar mixing angle.}

\subsection{Hierarchical masses}
Here we consider
\bea
M_\nu^h = \left[
                 \begin{array}{ccc}
                 0 & 0 & x_1  \\
                 0 & x_2 & x_3 \\
                 x_1 & x_3 & x_4
            \end{array} \right],
\label{a1}
\eea
then at the low scale it becomes 
\bea
M_\nu^l = a \left[
                 \begin{array}{ccc}
                 0 & 0 & {x_1}(1+\frac{r}{2})  \\
                 0 & x_2 & {x_3}(1+\frac{r}{2}) \\
                 {x_1}(1+\frac{r}{2}) & {x_3}(1+\frac{r}{2}) & {x_4}(1+r)
            \end{array} \right].
\nonumber
\eea
This is the type $A_1$ texture of \cite{fgm}.

After the rotation through $U_{23}$, 
\bea
{M_\nu^l}_{23} = a \left[
                 \begin{array}{ccc}
                 0 & -x_1 s^l_{23}(1+\frac{r}{2})
                 & x_1 c^l_{23}(1+\frac{r}{2}) \\
                 -x_1 s^l_{23}(1+\frac{r}{2}) & \lambda_1 & 0 \\
                 x_1 c^l_{23}(1+\frac{r}{2}) & 0 & \lambda_2
            \end{array} \right],
\nonumber
\eea
where
\be
\tan2\theta_{23}^l = \frac{2x_3(1+\frac{r}{2})}{x_4(1+r) - x_2}.
\label{a23}
\ee
To get
maximal mixing in the atmospheric sector through RG running, one must
therefore have
\be
x_4(1+r) = x_2.
\label{ares}
\ee
Further,
\be
\lambda_1 = x_2 c^{l2}_{23} + x_4 s^{l2}_{23}(1+r) - 2x_3 s^{l}_{23}
c^{l}_{23}
(1+\frac{r}{2}) \sim x_2 - x_3(1+\frac{r}{2}),
\label{al1}
\ee
\be
\lambda_2 = 
x_2 s^{l2}_{23} + x_4 c^{l2}_{23}(1+r) + 2x_3 s^{l}_{23} c^{l}_{23}
(1+\frac{r}{2}) \sim x_2 +
x_3(1+\frac{r}{2}),
\label{al2}
\ee
where (\ref{ares}) has been used in the second step.

The next rotation is through $U_{13}$ and is given by
\[
\tan2\theta_{13}^l = \frac{2x_1 c^l_{23}(1+\frac{r}{2})}
{\lambda_2}.
\]
Experiments indicate $\tan 2\theta^l_{13} \rightarrow 0$. Therefore, we
must have   $|\l_2| \gg |\sqrt{2}x_1(1+r/2)|$ which, in view of
(\ref{al2}), implies
\be
        |x_2 + x_3(1+\frac{r}{2})| \gg \sqrt{2}|x_1(1+\frac{r}{2})|.
\label{at1}
\ee
After this second rotation, the mass matrix ${M_\nu^l}_{23,13}$
looks like
\bea
{M_\nu^l}_{23,13} = a \left[
                 \begin{array}{ccc}
                 0 & -x_1 s^l_{23}(1+\frac{r}{2})
                 & 0\\
                 -x_1 s^l_{23}(1+\frac{r}{2}) & \lambda_1 & 0 \\
                 0 & 0 & \lambda_2
                 \end{array} \right].
\nonumber
\eea
For the remaining (12)-rotation, using (\ref{al1}), we have
\[
\tan2\theta_{12}^l   = \frac{-2x_1 s^l_{23}(1+\frac{r}{2})}
{x_2 - x_3(1+\frac{r}{2})},
\]
and the condition to get large $\tan2\theta_{12}^l$ is (using
$\sin\theta_{23}^l \sim 1/{\sqrt{2}}$)
\be
|x_2 - x_3(1+\frac{r}{2})| \ll \sqrt{2}|x_1(1+\frac{r}{2})|.
\label{at2}
\ee

Thus all the angles in the low energy scale will be compatible with the
data so long as the requirements in Eqs.~(\ref{ares}), (\ref{at1}),
(\ref{at2}) are met. Our aim now is to check whether these angles can
be generated from significantly different ones at the high scale.
Several alternatives are possible.

First, assume that $\theta_{23}$ is small at the high
scale. From (\ref{a23}), this implies
\[
2|x_3| \ll |x_4 - x_2| = r |x_4|,
\]
where we have used (\ref{ares}) in the last step. Since the
magnitude of $r$ $\sim 10^{-4}$, Eq.~(\ref{ares})
implies that $x_2 \simeq x_4$ and consequently $|x_2| \gg |x_3|$.
Thus, $|x_2 \pm x_3| \simeq |x_2|$ and hence conditions (\ref{at1}) and
(\ref{at2}) cannot be satisfied simultaneously. This establishes
that a radiative enhancement of $\theta_{23}$ is not
possible with large $\theta_{12}^l$ and small $\theta_{13}^l$ for
this mass matrix texture.

For the remainder of this subsection we restrict ourselves to the
situation where the angle $\theta_{23}$ does not
evolve much, i.e., $\theta_{23}^h \sim \theta_{23}^l = \p/4$. This
requires that
\be
|x_3| \gg |x_4 - x_2| =  r |x_4|.
\label{a4}
\ee
An evolution of $\tan 2\theta_{13}$ from a large value at the high
scale will occur if, in addition to
condition (\ref{at1}), one also has
\be
|x_2 + x_3| \ll \sqrt{2} |x_1|.
\label{a3}
\ee
Together, they require $|r x_3| \gg |x_1|$, which in view of the
smallness of $r$ implies $|x_3| \gg |x_1|$. From (\ref{a3}), then 
$x_2 \simeq -x_3$ and hence (\ref{at2}) cannot be satisfied. Thus a
significant running of $\theta_{13}$ is also excluded.

Finally, there is one remaining avenue for important RG evolution. This
is the case where $\tan 2\theta_{23}$ remains large over the entire
range while $\tan 2\theta_{13}$ is small, but $\tan 2\theta_{12}$
starts off small at the high scale and evolves to a near maximal
value at low scale.  We now show that even this is
inadmissible.  For the running of $\tan 2\theta_{12}$ one must
satisfy (\ref{at2}) as well as
\be
|x_2 - x_3| \gg \sqrt{2}|x_1|.
\label{a5}
\ee
These conditions can be simultaneously met if $|r x_3| \gg
|x_1|$. Therefore, $|x_3| \gg |x_1|$, which together with
(\ref{at2}),  (\ref{a4}) and (\ref{a5}) requires
\[
|x_4| \simeq |x_2| \simeq |x_3| \gg |x_1|/r.
\]
Here it is useful to note that this texture
corresponds to a hierarchical mass spectrum with eigenvalues
$\sim m, -m, M$ with $M \gg m$ \cite{dproy}. Therefore, $M \sim
\sqrt{\Delta m^2_{atm}} \sim 5 \times 10^{-2}$ eV. From the trace and
the determinant of the mass matrix (\ref{a1}) one finds
\[
M = a (x_2 + x_4) \simeq 2 a x_2 \;\;\; {\rm and} \;\;\;
M m^2 = a^3 x_2 x_1^2.
\]
Thus, $m \sim x_1 \ll r x_2 \sim 10^{-6}$ eV, which is way too small to
accommodate a possible solar mass square splitting at the level of
$10^{-5}$ eV$^2$ (LMA) or $10^{-7}$ eV$^2$ (LOW). 

A variant of this mass matrix texture -- type $A_2$ -- has the zero
off-diagonal entry in the (13) position. It is readily seen that this
does not affect the discussion concerning $\theta_{23}$ above. Since
this angle is maximal, after this rotation the $A_2$ texture and $A_1$
texture, considered above, give rise to identical structures and the
rest of the discussion goes through without change. {\em Thus, we draw
the conclusion that the hierarchical mass structures do not admit a
radiative enhancement of the mixing angles.}

\subsection{Inverted mass hierarchy}
Finally, we consider the neutrino mass matrix texture 
\bea
M_\nu^h = \left[
                 \begin{array}{ccc}
                 x_1 & x_2 & x_3  \\
                 x_2 & 0 & x_4 \\
                 x_3 & x_4 & 0
            \end{array} \right],
\nonumber
\eea
and therefore
\bea
M_\nu^l = a \left[
                 \begin{array}{ccc}
                 x_1 & x_2 & x_3(1+\frac{r}{2})  \\
                 x_2 & 0 & {x_4}(1+\frac{r}{2}) \\
                 {x_3}(1+\frac{r}{2}) & {x_4}(1+\frac{r}{2}) & 0
            \end{array} \right].
\nonumber
\eea
This texture (type $C$) results in a neutrino mass spectrum with an
inverted hierarchy \cite{fgm, dproy}.

We see that this structure with $(M_\nu)_{22} = (M_\nu)_{33} = 0$
ensures maximal mixing in the (23)-sector at all energies. Therefore,
\bea
{M_\nu^l}_{23} = a \left[
                 \begin{array}{ccc}
                 x_1 & (x_2 - x_3 (1+\frac{r}{2}))/\sqrt{2}
                 & (x_2 + x_3 (1+\frac{r}{2}))/\sqrt{2} \\
                 (x_2 - x_3 (1+\frac{r}{2}))/\sqrt{2}  & \lambda_1 &
                 0 \\
                 (x_2 + x_3 (1+\frac{r}{2}))/\sqrt{2} & 0 & \lambda_2
            \end{array} \right],
\nonumber
\eea
where
\[
\lambda_1 = -x_4(1+\frac{r}{2})\;\;\; {\rm and}\;\;\;
\lambda_2 = x_4(1+\frac{r}{2}).
\]
Now we have,
\be
\tan2\theta^l_{13} = \frac{\sqrt{2}[x_2 + x_3 (1+\frac{r}{2})]}
{x_4 (1+\frac{r}{2}) - x_1}.
\label{cc1}
\ee
In order to have a small $\theta^l_{13}$ as required by the CHOOZ
constraint, one must ensure 
\begin{equation}
|\sqrt{2}[x_3(1+\frac{r}{2}) + x_2]| \ll |x_4(1+\frac{r}{2}) - x_1|. 
\label{cc5}
\end{equation}
After the (13)-rotation, we then have
\bea
{M_\nu^l}_{23,13} = a \left[
                 \begin{array}{ccc}
                 x_1 & (x_2 - x_3(1+\frac{r}{2}))/\sqrt{2}
                 & 0 \\
                 (x_2 - x_3(1+\frac{r}{2}))/\sqrt{2} & \lambda_1 &
                 0 \\
                 0 & 0 & \lambda_2

                 \end{array} \right].
\label{2313}
\eea
Then diagonalizing the (12)-sector, we obtain 
\be
\tan2\theta^l_{12} \simeq 
\frac{\sqrt{2}[x_2 - x_3(1+\frac{r}{2})]}{\lambda_1 - x_1}. 
\ee
Now, we require a near  maximal mixing in the solar sector, which 
implies 
\begin{equation}
|\sqrt{2}[x_3(1+\frac{r}{2}) - x_2]| \gg |x_4(1+\frac{r}{2}) + x_1|.
\label{cc6}
\end{equation}
We should note that a simultaneous fulfillment of (\ref{cc5}) and
(\ref{cc6}), as dictated by experimental data, requires that
either one or both of (a) $x_1 \simeq - x_4$ and (b) $x_2 \simeq -
x_3$ have to be necessarily satisfied.

Using (\ref{cc5}) and (\ref{cc6}), we now calculate the solar and
atmospheric mass splittings from (\ref{2313}) as
\bea 
\Delta m^2_{sol} \simeq
\sqrt{2} a^2\left|[x_2 - x_3(1+\frac{r}{2})]
[x_1 - x_4(1+\frac{r}{2})]\right|, ~~~~~
\Delta m^2_{atm} \simeq  
0.5 a^2 \left[x_2 - x_3(1+\frac{r}{2})\right]^2.
\label{split}
\eea 

%Now $\Delta m^2_{sol} \ll \Delta m^2_{atm}$ implies 
%\be
%\left|x_1 - x_4(1+\frac{r}{2})\right| \ll 
%\left|x_2 - x_3(1+\frac{r}{2})\right|. 
%\label{cc7}  
%\ee

Now let us first study whether it is at all possible to accommodate the
running of $\theta_{13}$. To evolve this mixing angle from a large
value at a high scale to a small value at the low scale, one must
ensure the following high scale condition 
\be 
\sqrt{2}\left|x_2 + x_3\right| \gg \left|x_4 - x_1\right|. 
\label{cc8}
\ee 
But if we now ensure that (a) $x_1 \simeq - x_4$
and (b) $x_2 \simeq - x_3$, the two near-equalities emerging from
(\ref{cc5}) and (\ref{cc6}), are 
simultaneously
satisfied, it automatically follows that an enforcement of 
(\ref{cc8}) runs into contradiction with experimental data. 
To demonstrate this, first notice that 
the solar and
atmospheric mass splittings, in this case, take the simple form 
\bea 
\Delta m^2_{sol} \simeq
4\sqrt{2} a^2 x_3 x_4, ~~~~~
\Delta m^2_{atm} \simeq  
2 a^2 x_3^2.
\label{split-simple}
\eea 
From this we can infer that $|x_3| \sim 10^{-2}$ and $|x_4| \sim
10^{-3}$, as we have already found the parameter $a$ to be order
unity, to reproduce the experimental $\Delta m^2_{sol}$ and $\Delta
m^2_{atm}$. Incidentally, the values of $x_i$ so obtained correspond
to an inverted mass hierarchy.  Now let us relate $x_2$ and $x_3$ as $x_2
= -x_3 (1+\delta)$, where $\delta$ is some small
parameter. Substituting this relation in (\ref{cc5}) and (\ref{cc8}),
we observe that $\delta \simeq r/2 \sim 10^{-4}$ for a successful
running of $\theta_{13}$. From (\ref{cc8}) it then follows $|x_4| \ll
10^{-6}$, which is very different from the value $|x_4| \sim
10^{-3}$ obtained above directly from the mass splittings.  This
leads to the conclusion that a significant RG evolution of
$\theta_{13}$ is not possible. Through a somewhat more lengthy
chain of arguments, which we do not advance here, we can as well
demonstrate that even if any one of (a) $x_1 \simeq - x_4$ and
(b) $x_2 \simeq - x_3$ is satisfied, $\theta_{13}$ running is not
possible. Indeed, satisfying both (a) and (b), i.e., the case we
presented above, helps to simplify the illustration.

Next we consider the possibility of significant RG evolution of
the solar angle $\theta_{12}$, keeping $\theta_{13}$ small
throughout the scale.  The condition for a small $\theta_{12}$ at
the high scale is
\be 
\sqrt{2}\left|x_3 - x_2\right| \ll \left|x_4 + x_1\right|.
\label{cc9}
\ee
Now the question is whether (\ref{cc9}) can be ensured without
contradicting experimental data. First, one should observe that
this case is very similar to the case of $\theta_{13}$ running, and the
two cases can be handled in similar fashion. Proceeding in the same way
as we did for $\theta_{13}$, we can demonstrate that
$\theta_{12}$ running is not possible as well. 
In other words, the necessary compliance of any one or both of (a) $x_1
\simeq - x_4$ and (b) $x_2 \simeq - x_3$, in conjunction with the
experimental values of $\Delta m^2_{sol}$ and $\Delta m^2_{atm}$,
would lead to results which are inconsistent with (\ref{cc9}). 
{\em Thus the inverted
hierarchical neutrino mass matrix texture will also not support a
significant RG running of the mixing angles.}

\section{Non-standard models}
In the previous sections we have based our discussion on the RG
evolution of the neutrino masses within the framework of the standard
model. Essentially all these results can be extended to the case of
supersymmetry and models of extra dimensions. In both these scenarios,
$a$ and $r$, which capture the essence of the RG effect on neutrino
masses, satisfy $a \sim {\cal O}$(1) and $|r| \ll 1$. The conclusions
which we have drawn in the previous sections rest on these features
alone. Though the precise numerical values of the mass matrix
parameters will undergo appropriate modifications due to the changes
in $r$ and $a$, the basic conclusion that RG running of mixing angles
can be prominent in the quasi-degenerate case but not in the
hierarchical or inverted hierarchical alternatives will still continue
to hold. For the sake of completeness, we summarize the main new
ingredients of these two scenarios below.

\subsection{Supersymmetry}
In the Minimal Supersymmetric Standard Model, the RG equation
for $\kappa$ is slightly different from Eq.~(\ref{rgk})
\cite{babu1,chan1}.  There are two Higgs doublets, the quartic scalar
coupling $\lambda$ is determined by the gauge coupling, and
supersymmetric particles can appear in the internal lines of the
one-loop Feynman diagrams contributing to the different
beta-functions. We do not present the modified equation here. It is of
similar form but the coefficients are different. Suffice it to say
that
\[
r \simeq -(Y_\tau^2/8\pi^2) \ln(M_X/M_Z). 
\]
Note that $r$ is of opposite sign from the SM. It needs to be borne in
mind that in supersymmetry  $Y_\tau$ depends on $\tan \beta$ and can be
larger than in the standard model. In fact, $|r|$ can become as large
as $\sim 10^{-2}$ in this case.

\subsection{Extra Dimensions}
The RG evolution of neutrino masses in models with compact extra
dimensions has been examined in \cite{gau}. The main differences
from  the SM are that (a) the coupling constants evolve with
energy scale as a power law rather than logarithmically, and (b) the
higher scale, where the coupling constants unify, is not very
large $\sim {\cal O}$(10 TeV). If there are $\delta$ extra
dimensions and if the compactification radius is given by $\mu_0^{-1}$
then due to the power law running above $\mu_0 \sim$ 1 TeV:
\[
r \simeq \left(\frac{3Y_\tau^2}{16\pi^2}\right)
\frac{X_\delta}{\delta}\left[\left(\frac{M_X}{\mu_0}\right)^\delta - 1
\right]
\sim 10^{-4}, \;\;\; {\rm where}\;\;\; X_\delta =\frac{2
\pi^{\delta/2}}{\delta . \Gamma(\delta/2)}.
\]
$r$ is of the same order as in the SM in spite of power law evolution.
This is a consequence of the curtailed evolution range -- from $\mu_0
\sim$ 1 TeV to $M_X \sim$ 30 TeV.

\section {Conclusions}
Two of the three neutrino mixing angles are large -- a situation not
encountered in the quark sector. Radiative enhancement of neutrino
mixings could be a possible mechanism behind this. First, we have
considered the generation of neutrino Majorana masses and mixings via a
dimension-5 nonrenormalizable interaction. Then we have reviewed the
radiative mechanism for the two flavour neutrino mass matrix before
moving to three flavour cases. For the latter, we concentrate on
two-zero mass matrix textures. Radiative corrections of different
categories of such structures, namely quasi-degenerate and hierarchical
(normal and inverted), have been considered. We observe that in order
to maintain consistency with experimental data on masses and mixings,
the atmospheric and the CHOOZ angle cannot have appreciable running.
Only the solar angle can have a possibility to evolve from a low value at
a high scale to a large value at a low scale, and that too only
for the quasi-degenerate mass structures with one of the
neutrinos having opposite CP parity from the other two.   The
overall conclusions do not change for supersymmetric and
extra-dimensional scenarios.

\section*{Acknowledgements}
GB thanks LPT, Univ. de Paris XI, Orsay, for hospitality while a major
part of the work was being done.  AR has been supported in part by
C.S.I.R., India while AS enjoys a fellowship from U.G.C., India. They
(AR and AS) are grateful to the Abdus Salam International Centre for
Theoretical Physics, Italy, where the work was initiated.


\begin{thebibliography}{99}
\bibitem{3gen} M.C. Gonzalez-Garcia, M. Maltoni, C. Pe$\tilde {\rm
n}$a-Garay, J.W. Valle, Phys. Rev. D 63 (2001) 033005; G.L. Fogli,
G. Lettera, E. Lisi, A. Marrone, A. Palazzo, A. Rotunno, Phys. Rev. D
66 (2002) 093008; A. Bandyopadhyay, S. Choubey, S. Goswami, K. Kar,
Phys. Rev. D 65 (2002) 073031.

\bibitem{sun} SNO Collaboration, Q.R. Ahmed {\em et al.},
Phys. Rev. Lett. 89 (2002) 011301; {\em ibid.} 89 (2002) 011302;
SuperKamiokande Collaboration, S. Fukuda {\em et al.},
Phys. Rev. Lett. 86 (2001) 5656.

\bibitem{atm} SuperKamiokande Collaboration, T. Toshito, 
hep-ex/0105023 (to appear in the proceedings of 36th
Rencontres de Moriond on Electroweak Interactions and Unified
Theories, Les Arcs, France, 2001).


\bibitem{chz} CHOOZ collaboration, M. Appolonio {\em et al.}, Phys.
Lett. B 466 (1999) 415.

\bibitem{babu1} K.S. Babu, C.N. Leung, J. Pantaleone, Phys. Lett. B 
319 (1993) 191. 

\bibitem{chan1}  P.H. Chankowski, Z. Pluciennik, Phys. Lett. B 316 
(1993) 312. 

\bibitem{casas} J.A. Casas, J.R. Espinosa, A. Ibarra, I. Navarro, 
Nucl. Phys. B 573 (2000) 652.     

\bibitem{chan2} P.H. Chankowski, S. Pokorski, Int. J. Mod. Phys. A 17
(2002) 575; P.H.  Chankowski, W. Krolikowski, S. Pokorski,
Phys. Lett. B 473 (2000) 109.

\bibitem{balaji1} K.R.S. Balaji, R.N. Mohapatra, M.K. Parida,
E.A. Paschos, Phys. Rev. D 63 (2001) 113002.

\bibitem{balaji2} K.R.S. Balaji, A. Dighe, R.N. Mohapatra, M.K. Parida,
Phys. Lett. B 481 (2000) 33.

\bibitem{antusch} S. Antusch, J. Kersten, M. Lindner, M. Ratz, 
Phys. Lett. B 544 (2002) 1. 

\bibitem{antu2} S. Antusch, M. Ratz, JHEP 0211 (2002) 010.  

\bibitem{fgm} P. H. Frampton, S. L. Glashow, D. Marfatia, Phys.
 Lett. B 536 (2002) 79.

\bibitem{dproy} B.R. Desai, D.P. Roy, A.B. Vaucher,
hep-ph/0209035.

\bibitem{xing} Z.Z. Xing, Phys. Lett. B 530 (2002) 159. See also, 
Z.Z. Xing, Phys. Lett. B 539 (2002) 85.

\bibitem{rrr} P. Ramond, R.G. Roberts, G.G. Ross, Nucl. Phys. B 406
(1993) 19. 

\bibitem{an} S. Antusch, M. Drees, J. Kersten, M. Lindler, M.
Ratz, Phys. Lett. B 525 (2002) 130.

\bibitem{mtc} M. Machacek, M.T. Vaughn, Phys. Lett. B 103 (1981)
427; T.P. Cheng, E. Eichten,  L.F. Li, Phys. Rev. D 9 (1974)
2259; C.T. Hill, C.N. Leung, S. Rao, Nucl. Phys. B 262 (1985)
517.

\bibitem{db} H.V. Klapdor-Kleingrothaus, A. Dietz, H. L. Harney,
I.V. Krivosheina, Mod. Phys. Lett. A 16 (2001) 2409.

\bibitem{tri} J. Bonn {\em et al.}, Nucl. Phys. B (Proc. Suppl.) 91
(2001) 273; see also KATRIN Collaboration, A. Osipowicz {\em et al.},
hep-ex/0109033 (letter of intent for next generation Tritium beta
decay experiment).


\bibitem{fogli} P.C. de Holanda, A. Yu. Smirnov, hep-ph/0205241;
G.L. Fogli, E. Lisi, D. Montanino, A. Palazzo, Phys. Rev. D 64 (2001)
093007; J.N. Bahcall, M.C. Gonzalez-Garcia, C. Pe$\tilde {\rm n}$a-Garay,
JHEP 0108 (2001) 014; A. Bandopadhay, S.  Choubey, S. Goswami, K. Kar,
Phys. Lett. B 519 (2001) 83; P.I.  Krastev, A. Yu. Smirnov,
Phys. Rev. D 65 (2001) 073022.


\bibitem{mns} Z. Maki, M. Nakagawa, S. Sakata, Prog. Theor. Phys. 28
(1962) 870.

\bibitem{galaxy} O. Elgaroy et al., Phys. Rev. Lett. 89 (2002) 061301.  

\bibitem{balaji3} K.R.S. Balaji, A. Dighe, R.N. Mohapatra, M.K. Parida,
Phys. Rev. Lett. 84 (2000) 5034. 

\bibitem{gau} G. Bhattacharyya, S. Goswami, A. Raychaudhuri,
Phys. Rev. D 66 (2002) 033008.

\end{thebibliography}
\end{document}